\documentclass[twocolumn,preprintnumbers,amssymb,amsmath,superscriptaddress,letterpaper]{revtex4}[10pt]

\usepackage{graphicx}
\usepackage{dcolumn}
\usepackage{bm}
\usepackage{natbib}
\usepackage{epsfig}
\usepackage{units}

\newcommand{\be}{\begin{equation}}
\newcommand{\ee}{\end{equation}}
\newcommand{\bea}{\begin{eqnarray}}
\newcommand{\eea}{\end{eqnarray}}

\newcommand{\sigmav}{\langle\sigma_Av\rangle}

\begin{document}

\hspace*{130mm}{\large \tt FERMILAB-PUB-13-080-A-PPD}
\vskip 0.2in

\title{The 111 and 129 GeV $\gamma$-ray lines from annihilations in the Milky Way \\ dark matter halo, dark disk and subhalos}

\author{Ilias Cholis}
\email{cholis@fnal.gov}
\affiliation{FNAL, Theoretical Astrophysics Group, Batavia, Illinois, 60510, USA}
\author{Hani Nurbiantoro Santosa}
\email{santosa@sissa.it}
\affiliation{SISSA, Via Bonomea, 265, 34136 Trieste, Italy}
\affiliation{INFN, Sezione di Trieste, Via Bonomea 265, 34136 Trieste, Italy}
\author{Maryam Tavakoli}
\email{maryam.tavakoli@desy.de}
\affiliation{II. Institut f\"ur Theoretische Physik,  Universit\"at Hamburg, Luruper Chaussee 149, 22761 Hamburg, Germany}
\author{Piero Ullio}
\email{ullio@sissa.it}
\affiliation{SISSA, Via Bonomea, 265, 34136 Trieste, Italy}
\affiliation{INFN, Sezione di Trieste, Via Bonomea 265, 34136 Trieste, Italy}
\date{\today}

\begin{abstract}
Recently a series of indications have been put forward suggesting the  presence of two $\gamma$-ray lines at 110-130 GeV (centered at 111 and  
129 GeV).  Signals of these lines have been observed toward the Galactic center, at some galaxy clusters and among some of the unassociated 
point sources of the 2 years Fermi  catalogue. Such a combination of signals could be generated by dark  
matter annihilations in the main dark matter halo, its substructures  and nearby galaxy clusters. We discuss in this work the consistency between  
the number of events observed at the line energies in the sky and the  predictions using results from the Via Lactea II numerical simulation  
and extrapolations below its mass resolution, taking into account that  the annihilation cross-section to the lines can be estimated from the  
Galactic center signal. We find that some extrapolations to small  substructures can naturally account for the point sources signal, although the  
hypothesis of background only cannot be rejected. We also study the  morphology of the $\gamma$-ray sky at the 2 lines energies, testing  
different Galactic diffuse background models to account for interstellar medium uncertainties and different assumptions on the DM  
diffuse component profile.
We find from template fits that within reasonable diffuse background  uncertainties the presence of a spherical halo component is preferred
with cuspier dark matter halo profiles being preferable even from the  full sky fit. We finally check the impact of a dark disk component  
suggested by cosmological simulations that include baryons and find that thin dark disks can not be disfavored, thus possibly accounting for the  
preferentially closer to the Galactic disk distribution of the point sources lines signal.
\end{abstract}

\maketitle

\section{Introduction}
\label{sec:intro}

The possible identification by the  \textit{Fermi} Gamma-ray Space Telescope of a signal compatible 
with the monochromatic photon emission due to pair annihilations of cold dark matter (DM) 
particles has recently been one of the most debated topics. Originally, 
\cite{Bringmann:2012vr, Weniger:2012tx} suggested the detection of a line at 
$129.8 \pm 2.4 ^{+7} _{-13}$ GeV with a 3.3$\sigma$ significance \cite{Weniger:2012tx} in a wide window toward the Galactic center (GC).  
A similar signal has been indicated by \cite{Su:2012ft} at $127.0 \pm 2.0$ GeV with a 5.0$\sigma$ significance.
A pair of lines with energies of $110.8 \pm 4.4$ and $128.8 \pm 2.7$ GeV can alternatively explain 
 $\gamma$-ray excess at 5.4$\sigma$ significance \cite{Su:2012ft}.  
Similarly the line signal at $\simeq$ 130 GeV has also been found by \cite{Tempel:2012ey} 
with \cite{Rajaraman:2012db, Buchmuller:2012rc, Boyarsky:2012ca} suggesting the presence of 2 lines at $\simeq$ 110 and 130 GeV.
Both line signals are in agreement with constraints from line searches of the \textit{Fermi} collaboration
\cite{Ackermann:2012qk} and indicate a preference for dark matter (DM) annihilation rather than decay 
\cite{Weniger:2012tx, Su:2012ft, Kuhlen:2012qw, Rao:2012fh} (for a recent review on DM line searches read \cite{Bringmann:2012ez}).
The \textit{Fermi} collaboration, motivated by the results of 
\cite{Bringmann:2012vr, Weniger:2012tx, Su:2012ft, Tempel:2012ey, Rajaraman:2012db, Buchmuller:2012rc, Boyarsky:2012ca}, 
has performed an analysis oriented toward the GC and some of those results are presented in 
\cite{Charles, Albert}. 
Similar to \cite{Finkbeiner:2012ez, Whiteson:2012hr}, no obvious systematic error has been found to account for the 
amplitude of the line signal measured by \cite{Weniger:2012tx, Su:2012ft} (see though \cite{Whiteson:2013cs}).
The \textit{Fermi} results confirm a line-like signal at E$\simeq$130 GeV at $4\sigma$, or E$\simeq$135 GeV at 3.3$\sigma$ significance after reprocessing the data to take into account the shift  of the reconstructed energy with time. 
Yet, some part of the amplitude may be related to limb photons \cite{Charles}.
Adding information on the performance of the instrument's energy reconstruction decreases the significance of the signal \cite{Albert}. 
Thus a conclusive answer on whether the line signal is a systematic error identified toward the GC or a signal of DM annihilations has not been provided yet. 

A monochromatic gamma-ray flux is expected in most scenarios in which DM is in the form 
of weakly interacting massive particles (WIMPs) since two-body annihilation final states containing
photons arise at the 1-loop level. At the same time, in such a framework, it is foreseen that
tree-level WIMP annihilations into other SM final states,  in turn hadronizing  and/or decaying 
into $p$, $\bar{p}$ $e^{\pm}$, $\nu$s and $\gamma$s, would give sizable yields with continuum
energy spectrum on top of the monochromatic $\gamma$-ray yield.
Yet, no clear indication of $\gamma$-ray excess, other than the lines, has been found toward the inner few degrees of the GC, 
leading to the extraction of constraints on the continuum \cite{Buchmuller:2012rc, Buckley:2012ws, Cohen:2012me, Cholis:2012fb}; and thus motivating 
further discussions for the particle physics origin of the lines \cite{Buckley:2012ws, Weiner:2012gm, Fan:2012gr}. 

Another aspect of a WIMP annihilation signal is that one should expect to see line signals at the same energy and annihilation cross-section
toward other dark matter targets. 
In \cite{Hektor:2012kc}, a 130 GeV line signal toward known galaxy clusters has been suggested. 
There have been also indications for two lines at 111 and 129 GeV in unassociated point sources which would imply DM annihilating in substructures 
\cite{Su:2012zg, Hektor:2012jc} 
\footnote{There have been16 unassociated point sources in the 2yr \textit{Fermi} catalogue detected 
at the level of $\simeq$4$\sigma$ or more. In oder to reach that significance a spectrum at low energies is also needed. 
DM does not give a strong spectrum signal at low energies so the actual spectrum
of these sources (if they are DM subhalos) may be contaminated by other close-by 
point sources or the diffuse $\gamma$-ray component. That may also result in a non-trivial selection effect.} (see also \cite{Hooper:2012qc} for an alternative interpretation). 

We assume that the line signal from the unassociated point sources is indeed of DM origin, with the same annihilation cross-section to the lines as is estimated from the GC.
We then confront that signal with predictions from cosmological simulations such as Via Lactea \cite{Diemand:2006ik, Diemand:2007qr, Kuhlen:2008aw}.
In section~\ref{sec:Assumptions}, we discuss the $\gamma$-ray data that we use and
our general assumptions for the background and the DM density distribution in the Galaxy.
As a general reference we use the total number of $\gamma$-ray events with energy $111 \pm 5$ GeV and
$129 \pm 6$ GeV. These events are taken to be of both DM and diffuse/point source $\gamma$-ray background origin.
The comparison of the observed $\gamma$-ray data with the
predicted contribution from substructures within the Galaxy's virial radius is done in section~\ref{sec:VLII}.
Our aim is to conclude on whether the line signal at the point sources can be physically associated to the same energy line 
signal toward the GC and under what assumptions on the substructure distribution. 
On the CDM simulation side we use VLII subhalo distribution data \cite{Kuhlen:2008aw}
 and also extrapolate the VLII simulation mass function to smaller subhalo
masses.

The observed non-isotropic distribution in the sky of the DM line(s) signal associated to the point sources, 
could be explained by the presence of a strong dark disk.
Such a dark disk would also have an impact on the diffuse distribution of the DM originated line photons. 
Independently, in the context of self-interacting DM, the formation of dark disks has been suggested to explain the relatively large amplitude 
of the line signal in the inner kpc of the Galaxy \cite{Fan:2013yva, Fan:2013tia}.  
To study the diffuse $\gamma$-ray sky at the energies of the 2 lines, one needs predictions on both the Galactic diffuse backgrounds and on the
DM diffuse contribution. 
In section~\ref{sec:LineEmissions}, we test that possible contribution to the two $\gamma$-ray lines
from DM annihilations in the Galaxy and from the diffuse backgrounds by doing a template fit. The importance of template fits is  that one can 
take into account the different morphologies of the various diffuse components. We study the impact on the significance of a DM signal on the 4$\pi$
sky (including the GC) of different assumptions
for the Galactic diffuse background, related to physical properties of the interstellar medium. 
We also test different assumptions (and thus different templates) on the main spherical halo density profile, on the significance of the dark disk
component to the local dark matter density and its thickness and finally on the contribution of the dimmer DM subhalos that would
also add to the diffuse $\gamma$-ray sky flux. We consider that the main contribution from the brighter DM structures
has been already observed by \cite{Su:2012zg} and exclude them from the $\gamma$-ray fits.
We also derive upper limits on the diffuse emission from annihilations in the main DM halo,
and give our conclusions in section~\ref{sec:Conclusions}.
\section{Gamma-Ray data, Diffuse Background and DM distribution Assumptions}
\label{sec:Assumptions}

The \textit{Fermi} Large Area Telescope publicly available events are categorized in different classes
based on the expected level of cosmic ray (CR) contamination. In this work we  use the
ULTRACLEAN events class which is the cleanest $\gamma$-ray events sample.
There are 686(744) photons with energy between $111 \pm 5$ GeV and 611(668)  photons
with energy between $129 \pm 6$ GeV in the 4 yr (4.4 yr) full sky \textit{Fermi} ULTRACLEAN class $\gamma$-ray data, with the
quoted energy ranges representing the relevant energy dispersion for these lines
 \footnote{Our 4.4 yr sample refers to the data from 04 August 2008-03 January  2013.}.
In section~\ref{sec:VLII} where we compare with the findings of \cite{Su:2012zg}, we use the 4yr sample since it approximates their events sample,
while in section~\ref{sec:LineEmissions} we use the slightly longer period of 4.4 yr .

The emission of the diffuse Galactic $\gamma$-ray background above 100 GeV is dominated by the $\pi^{0}$ contribution, i.e. the decay of mesons produced by
inelastic collisions of CRs with the interstellar medium (ISM) gas and by the up-scattering of low energy photons of the interstellar radiation field (ISRF) from high energy CR electrons (inverse Compton scattering). 
The morphology of these components on the sky is different mainly because of the different distributions of the ISM gas density and the ISRF energy density in the Galaxy.
Moreover the energy loss of CR electrons and protons during their propagation in the Galaxy is different.
The bremsstrahlung radiation off CR electrons at these energies is completely subdominant but is included in our code.

To compute the diffuse $\gamma$-ray background, we use the DRAGON package \cite{Evoli:2008dv, DRAGONweb, Cholis:2011un} with a new ISM gas model \cite{Tavakoli:2012jx} that ensures good agreement with $\gamma$-ray spectral data between 1 and 200 GeV in the full sky and subsections of it \cite{MaryamEtAl}. 
We ignore the contribution of the "dark gas" (not related to DM substructures) whose uncertainties are significant in the inner $5^{\circ}$ in latitude
\cite{2005Sci...307.1292G, FermiLAT:2012aa}.
Based on the relevant uncertainties (see \cite{MaryamEtAl}), we allow for
different assumptions on the ISM gas and the ISRF which influence the $\pi^{0}$ and the inverse Compton $\gamma$-ray 
emissivities respectively.
   
In the case where there are 2 lines as has been indicated by \cite{Rajaraman:2012db, Su:2012ft}, the energy of these lines is centered at 
$128.8 \pm 2.7$ and $110.8 \pm 4.4.$ GeV \cite{Su:2012ft}.
The lines come from either the combination of $2\gamma \& Z \gamma$ lines or from the $Z \gamma \& h \gamma$ lines.    
   
In \cite{Cholis:2012fb}, five individual modes/channels of DM annihilation:
$\chi \chi \longrightarrow W^{+}W^{-}$, $\chi \chi \longrightarrow b\bar{b}$,
$\chi \chi \longrightarrow \tau^{+}\tau^{-}$, 
$\chi \chi \longrightarrow \mu^{+}\mu^{-}$ and 
$\chi \chi \longrightarrow e^{+}e^{-}$ have been studied.
The limits on the DM annihilation cross-sections based on their contribution to the continuum $\gamma$-rays spectrum in the $\mid l \mid < 5^{\circ}$, $\mid b \mid < 5^{\circ}$ observation window have been derived. 
Typically, DM models have sizable branching ratios into more than one of these channels.
Yet apart from the $\chi \chi \longrightarrow \mu^{+}\mu^{-}$ channel and mainly the
$\chi \chi \longrightarrow e^{+}e^{-}$ channel, in all the other annihilation channels to SM 
particles with a continuum spectrum, the $\gamma$-ray DM signal at 111 and 129 GeV can not
be explained/mimicked by the continuum spectrum. 
Thus it originates from the annihilation into $Z\gamma$ and $2 \gamma$. For $\chi \chi \longrightarrow e^{+}e^{-}$
and $\chi \chi \longrightarrow \mu^{+}\mu^{-}$, the final state radiation (FSR) and virtual
internal bremsstrahlung (VIB) can contribute to the line signal as discussed in 
\cite{Bringmann:2012vr, Cholis:2012fb}. 

For simplicity we assume that the DM induced $\gamma$-rays with energy $111 \pm 5$ GeV and $129 \pm 6$ GeV come from
the annihilation of a 129 GeV DM particle into $Z \gamma$ and $2 \gamma$ respectively.
Alternatively, these $\gamma$-ray lines could come from $h \gamma$ and $Z \gamma$ for the case of a 142 GeV DM particle.  
The relevant ratio of the luminosity of two lines is taken to be 1/2 for the 111/129 GeV lines as suggested in \cite{Su:2012zg}, thus for the case of 129 GeV DM particle, the annihilation cross-sections to $Z \gamma$ and $2 \gamma$ are assumed to be the same.

For the DM distribution we assume that it is a combination of a spherically symmetric "main" 
DM halo and a dark disk (DD). For the main halo we assume a spherical  Einasto DM profile:
\begin{equation}
\rho_{sph}(r) = \rho_{Ein} \exp\left\{-\frac{2}{\delta}\left[\left(\frac{r}{r_{c}}\right)^{\delta}-1\right]  \right\},
 \label{eq:Einasto}
\end{equation}
using $\delta = 0.13, 0.17, 0.22$ \cite{Merritt:2005} with $r_{c} = 20$ kpc.
The values of $\delta =0.13 (0.22)$ result in a more (less) cuspy DM distribution. 
The density normalization parameter 
$\rho_{Ein}$ is set in terms of the local DM density, after including a contribution of the DD.

The profile of the DD component is assumed to be described by \cite{Read:2008fh}:
\begin{equation}
\label{eq:DD_eq}
\rho_{DD}(R,z)=\rho_{0_{DD}} \exp\left[\frac{1.68 \left(R_{\odot}-R\right)}{R_{1/2}}\right]\exp\left[-\frac{0.693\left|z\right|}{z_{1/2}}\right],
\end{equation}
where $R_{1/2}$ and $z_{1/2}$ are the half mass scale lengths in the Galactic plane and perpendicular to the Galactic plane, respectively 
and $R_{\odot}=8.5$ kpc.  
Here $R$ is the cylindrical radial coordinate. 

The ratio of the local DM density in the dark disk to the local DM density in the 
spherical halo $\rho_{0_{DD}}/\rho_{0_{sph}}$ typically ranges between 0.2-1.5 
\cite{Read:2008fh}, with the higher ratios being related to higher mass densities in 
the thick stellar disk rather than in the thin stellar disk. The thick stellar disk
can be populated by thin stellar disk stars, if the thin stellar disk gets heated by very 
massive, high-redshift mergers.  Another cause could be multiple pro-grate and 
low inclination mergers \cite{Read:2008fh}. 

In the template analysis performed below we will restrict to the case:
\begin{equation}
  \label{eq:init_contraints2}
  \alpha/2 \equiv  \rho_{0_{DD}}/(\rho_{0_{sph}}+\rho_{0_{DD}}) \leq 0.5.
\end{equation}
fixing \cite{Catena:2009mf,Salucci:2010qr}:
\begin{equation}
  \label{eq:init_contraints}
  \rho_{0_{sph}} + \rho_{0_{DD}} = 0.4\;\textrm{GeV cm}^{-3}.
\end{equation}  
Regarding the dark disc thickness, 
some authors \cite{Kalberla:2007sr} have suggested thicker disks, while thinner and 
less significant dark disks can also be the case. Keeping in the parametrization of eq.~\ref{eq:DD_eq}
$R_{1/2} = 11.7$  fixed \cite{Read:2008fh}, we will test the half mass scale length values of 
$z_{1/2} =$ 0.5, 1.0, 1.5, 3.0 kpc. 

In the standard model for cosmology, cold DM structures form hierarchically, with small 
DM halos collapsing first and subsequently merging into larger and larger objects. Since
tidal disruption may only be partially effective, massive DM halos, such as the 
halo of our own Galaxy, are expected to contain a vast population of subhalos, with mass spanning 
from a tiny seed mass up to a fraction of the hosting halo mass. The minimum mass is 
essentially associated to the free-streaming scale of DM particles, in turn depending on
their temperature of kinetic decoupling in the early Universe. For WIMPs the minimum mass 
can be as small as about $m_{cut}=10^{-6}M_{\odot}$ \cite{Green:2003un,Profumo:2006bv},
much lighter than the dwarf galaxy scale, possibly to the smallest environment which can 
host stellar populations and hence a luminous counterpart. Because of the highly non-linear 
nature of the merging process,  up to now the only efficient technique to model in detail DM 
halos is the use of numerical N-body simulations with large populations of substructures 
 found in such studies. We will assume as primary reference in our analysis 
results from Via Lactea II (VLII)\cite{Diemand:2008in}, one of the
highest resolution simulations up to date of a Milky Way-sized CDM halo (virial mass
$M_h=1.9\times10^{12}M_{\odot}$), with over one billion DM "particles" and
nominal mass resolution of about  4100 $M_{\odot}$ (numerical effects appear to enter
well above this scale, possibly affecting the subhalo mass spectrum up to 
about $\sim3\times10^{6}M_{\odot}$).  In our analysis we will discuss  both the DM
pair annihilation associated to individual DM substructures as well as the collective 
effect from the whole subhalo population. In both respects, the resolution of the simulations 
appears insufficient to properly model the expected signals. Our approach will then be 
to use the simulation results to properly calibrate the necessary extrapolations to smaller 
masses: tuning, at a given Galactocentric radius, the subhalo pericenter distribution and 
applying a recipe for taking into account tidal stripping effects. We derive a model 
which reproduces fairly well the subhalo mass function and the distribution in halo concentration 
as a function of radius in the VLII simulation, and we use it as a prediction below 
its resolution (some details about our approach are given in Appendix~\ref{sec:UnBiasedDistr}).

\begin{figure}
\begin{centering}
\includegraphics[bb=1.3cm 6.5cm 19.5cm 21.5cm,clip,scale=0.4]{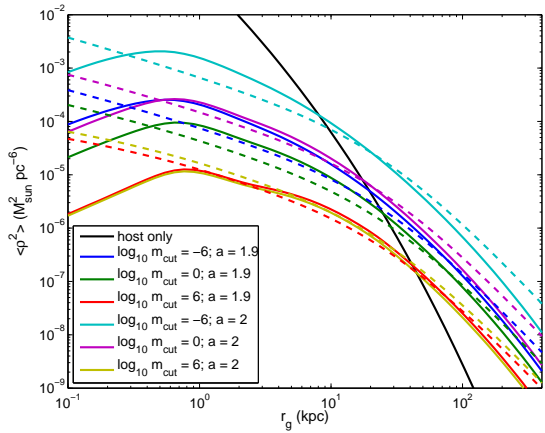}
\par\end{centering}
\caption{\label{fig:rhosbsqrd} Average DM squared density $\langle\rho^2\rangle$ 
associated to subhalo population as a function of Galactocentric radius $r_g$, 
for the case of an unbiased distribution (solid lines) and an anti-biased distribution 
(dashed lines) and for a few assumptions on the values for subhalo mass function 
$a$ and minimum subhalo mass $m_{cut}$ (see the text for details). Also shown 
is the density squared of the smooth halo component.}
\label{fig:rho2}
\end{figure}

The general trends in the DM subhalo distribution can be understood from the fact that
more massive objects are more prone to tidal stripping than the less massive ones, because 
they typically have smaller average density, reflecting the fact that they collapsed later 
in the cosmic history at a lower averaged background density. As a result, when going 
toward the center of the host halo, the average subhalo density increases and the average 
mass decreases. Also, in the inner part of a DM halo the tidal forces become stronger,
possibly making the subhalos spatial distribution to be anti-biased with respect to host 
density profile. This tendency has been found in some numerical simulations, see, 
for example~\cite{Kuhlen:2008aw,Springel:2008cc}, while it has also been claimed that 
unbiased or anti-biased distributions may just stem from selection effects \cite{Diemand:2007qr,Diemand:2008in,Diemand:2009bm,Anderson:2010df}. 
The model in our extrapolation from the VLII results gives an
unbiased distribution. To bracket uncertainties 
we consider also an anti-biased distribution which is instead taken from \cite{Kuhlen:2008aw}. 
The other parameters entering most critically in our analysis are the 
spectral index $a$ for the subhalo mass function and the minimum subhalo mass 
$m_{cut}$. In Fig.~\ref{fig:rho2}, we plot as a function of Galactocentric radius, the 
average DM squared density associated to the full subhalo population for two choices
of the spectral index, i.e. $a=1.9$ and $a=2$, three sample values of the minimum subhalo 
mass, and for the unbiased (solid lines) and anti-biased (dashed lines) distributions. For a 
comparison the density squared of the smooth DM halo component, 
Eq.~\ref{eq:Einasto}, with $\delta=0.17$, is also given.

\section{DM substructures in the Milky Way and the 2 $\gamma$-ray lines}
\label{sec:VLII}
In \cite{Su:2012zg}, 16 unassociated point sources have been identified with at least one 100-140 GeV photon for SOURCE event class .
Among those photons, there are 9 photons in the ULTRACLEAN event class sample with energies $\sim$111 and $\sim$129 GeV.
Those photons from unassociated point sources may imply a signal of DM annihilation in close by substructures. 
That ansatz can be compared to predictions from cosmological simulations.
 
The 2 yr point source catalogue \cite{Fermi:2011bm} has 575 unassociated point sources. 
In \cite{Su:2012zg}, the 9 ULTRACLEAN photons at $\simeq$ 111 and 129 GeV
lie within a $0.15^{\circ}$/$0.3^{\circ}$ radius for FRONT/BACK converted events covering an area of 0.07/0.28 square degrees.
Thus the 575 unassociated point sources cover at most $1.6 \times 10^{2}$ deg2 out of $4.1 \times 10^{4}$ deg2 and would
give a conservative upper estimate of 5.06 photons out of the $1.3 \times 10^3$ 
\footnote{This value is obtained assuming isotropic distribution. The actual distribution of the 111 and 129 GeV photons is not isotropic but rather peaks on the Galactic disk. Our results do not differ that much though since the distribution of the 575 unassociated point sources also peaks on the Galactic disk.}.
The probability that 9 or more ULTRACLEAN class photons out of the $1.3 \times 10^3$ photons of the 4$\pi$ sky fall within the area
 covered by the 575 unassociated point sources is $p=0.0721$ 
 \footnote{\cite{Su:2012zg} suggest a p-value of $p=0.00069$ for having 13 SOURCE events given a background described by a power-law.}. 
Yet there may also be a bias toward discovering point sources around single high energy events \cite{Alexprivate}. 
While by itself not a strong statistical deviation, the coincidence in energy with the GC signal inclines us to test for alternative possibilities and whether some of these photons come from unassociated point sources that are DM subhalos.
    
The number of photons that we receive from a single subhalo with luminosity $L \equiv \int\rho_{sub}^2\mathrm{d}V$ and line of sight (los) distance $\lambda$ from us, for channel $ch=\gamma\gamma$ or $\gamma Z$ is given by:
\begin{equation}
N^{ch}=\mathcal{N}_{\gamma}^{ch}\frac{\sigmav_{ch}}{2}\frac{L}{m_{\chi}^{2}}\frac{\tau_{exp}A_{exp}}{4\pi \lambda^{2}}\;,\label{eq:numgam}
\end{equation}
 where $m_{\chi}=\unit[129]{GeV}$
is the DM particle mass and $\mathcal{N}_{\gamma}^{ch}=1(2)$ for
$\gamma Z\left(\gamma\gamma\right)$. $\tau_{exp}$ and $A_{exp}$
are the detector's exposure time and effective area for photon's energy
of 129 GeV, respectively. 
In this work, we use for the averaged \textit{Fermi}-LAT exposure after $\simeq4$ years
$\tau_{exp}\times A_{exp}=\unit[1.22\times10^{11}]{cm^{2}s}$.  
For this section, we consider two values of annihilation rates: $\langle \sigma v \rangle_{\gamma\gamma} = $ 
$\langle \sigma v \rangle_{Z\gamma} = 0.98 \times 10^{-27}$ cm$^3$ s$^{-1}$,
a value derived assuming our default smooth component DM density profile (no DD) and fitting 
the monochromatic signal in the region $\mid l \mid < 5^{\circ}$ \& $\mid b \mid < 5^{\circ}$~\cite{Cholis:2012fb}; 
and $\langle \sigma v \rangle_{\gamma\gamma} = \langle \sigma v \rangle_{Z\gamma} = 3 \times 10^{-28}$ cm$^3$ s$^{-1}$, which fits better the whole sky region 
(see section~\ref{sec:LineEmissions}).
   
To quantify the possible impact of substructures in the line photons on the sky we ask the following questions:\\
A) How many subhalos give 2 or more photons in to the 2 $\gamma$-ray lines?\\
B) How many photons (in the two lines energies) do we get from all the subhalos that 
give a more than 0.1 photons?\\
The difference between the number of photons from the entire subhalo population received and the answer to question (B) is a proxy for the 
 diffuse gamma-ray flux to the two lines from DM substructures gravitationally bound 
in the main DM halo. We will refer to these photons as "DM substructure diffuse". 

In Table~\ref{tab:SubhaloContribution} we consider first the single subhalo sample from the
VLII simulation~\cite{Diemand:2008in} (no extrapolation below the mass resolution at this level)
and compute answers to the questions formulated above, averaging over results obtained for 100 random choices for the position of the observer, all at fixed Galactocentric distance $R_{\odot}=\unit[8.5]{kpc}$ (the average is performed to wipe out fluctuations involving effects of nearby subhalos  or voids in a single random choice). We then turn our analysis to the extrapolated subhalo populations
focussing on the unbiased and anti-biased distribution and considering two possible extrapolations for lowest subhalo mass and subhalo mass spectral index, bracketing extreme possibilities, 
case I corresponding to $(m_{cut},a)=(10^3\,{\rm M}_\odot, 1.9)$ and 
case II to $(m_{cut},a)=(10^{-6}\,{\rm M}_\odot, 2)$. 
The number of line photons received is computed including only subhalos within the virial
radius since we find that for VLII subhalos only $0.5 \%$ of the total photon comes from substructures
lying outside it. On the other hand there is a further uncertainty one should be careful about: 
by tuning our subhalo model to the VLII  results we are fixing the normalization of subhalo 
number density (above mass resolution) for the Milky Way halo according to that specific 
realization. This is a quantity which actually
has a certain scattering among different halo realizations and different simulations. 
For instance if we wanted to follow the results of the Aquarius simulation~\cite{Springel:2008cc} we should increase
the normalization of the subhalo number density by about a factor of 2, shifting results in Table~\ref{tab:SubhaloContribution}
by the same factor.

As a further test, in Table~\ref{tab:SubhaloDistribution} we report the total number of monochromatic 
photons expected from our entire subhalo population under a set of different assumptions for spectral index and cutoff mass. The results are again shown for unbiased and anti-biased distributions. 

\begin{table}[t]
\begin{tabular}{|c||c||c|}
\hline
Simulation Assump.	& Q.A			& Q.B	 \\
\hline \hline
VLII			& 0 (0)    		& 0.213 (0.024)	 \\
\hline
 unbiased - case I	& 0.0198 (0.00344)		& 0.473 (0.0874) \\
\hline
 unbiased - case II	& 0.0139 (0.0024)	 	& 0.342 (0.0618)  \\
\hline
anti-biased - case I  	& 0.0746 (0.0176)		& 1.24 (0.296)	\\
\hline
anti-biased - case II 	& 0.0898 (0.0196)		& 1.62 (0.361)	\\
\hline 
\end{tabular}
\caption{Relevance of substructures for detection of the monochromatic photons referring to
 questions A and B as posed in the text, using cross section which fits GC (fits the whole sky).
Answers are provided in case of the subhalo sample 
from the VLII simulation itself and from extrapolations of it in case of
unbiased or anti-biased distributions with the parameter choice 
$m_{cut}=10^3\,{\rm M}_\odot$ $\&$ $a=1.9$ (case~I) and  
$m_{cut}=10^{-6}\,{\rm M}_\odot$ $\&$ $a=2$ (case~II). Changing the overall 
normalization of the subhalo number density would shift the results provided in the table
accordingly, e.g., by a factor of about 2 if adopting the normalization of the Aquarius simulation~\cite{Springel:2008cc}.}
\label{tab:SubhaloContribution}
\end{table}

If the photons from unassociated point sources are from DM annihilation in substructures, their number (9 ULTRACLEAN events) 
probes the number of photons  from the brighter substructure subsample. 
Considering that the number of photons originating from subhalos that emit more than 0.1 photons in the two lines 
(i.e. the results of Question B in Table~\ref{tab:SubhaloContribution}), is indicative of such number of photons, we compare the probability of having 
observed 9 (or more) photons for a DM signal calculated in the VLII sample (0.213 photons), in extrapolation 
for unbiased distribution and case I (0.473 photons), 
for anti-biased distribution and case I (1.24 photons), 
for unbiased distribution and case II  (0.342 photons), and 
for anti-biased distribution and case II (1.62 photons). 
These probabilities are $p=0.0874$ for the background plus the DM signal stemming from VLII sample, $p=0.108 (0.185)$  for the background plus the DM signal in the extrapolation for unbiased (anti-biased) distribution in case I and $p=0.0975(0.23)$ for background plus DM signal in the extrapolation for unbiased (anti-biased) distribution in case II.
Using the normalization from Aquarius simulation would increase the number of subhalos and received photons by a factor of 2, shifting the probabilities
to, respectively,  $p=0.153(0.344)$ and $p=0.127(0.449)$.
Thus the most conservative VLII assumptions case is marginally favorable than the just background case.
When extrapolating below the mass resolution, probabilities increase
further, reaching relevant levels in optimistic extrapolations.
On the other hand, using the cross section which fits the whole sky (see section~\ref{sec:LineEmissions}), we don't see much differences in the p-values from having just a background signal. Also, from Question A, there are no subhalos expected to give more than 2 photons. One must keep in mind that lowering the value of cross section by a factor of 3, by going from $\sigma v$ which fits GC better to the one which fits the whole sky better, does not simply reduce all the values in Table I by the same factor. This is because the photons produced by each subhalo will decrease; so that some subhalos which previously gave more than, say, 0.1 photons; will now give less. The photons coming from such subhalos are not included anymore. (Similarly with the numbers of subhalos which give more than 2 photons.) However, in Table II, all values do lower by the same factor, because they are the number of photons coming from all subhalos.
 
The differences between the numbers of photons that originate from all DM subhalos (Table~\ref{tab:SubhaloDistribution}) and the numbers 
of photons that originate from DM subhalos that contribute $10^{-1}$ lines photons or more (Q.B) are conservative probes 
to the diffuse contribution from the DM subhalos at $\simeq$ 111 and 129 GeV. 
The VLII sample gives $1.21-0.213 = 1.0$ DM substructure diffuse component photon, 
extrapolation for the unbiased (anti-biased) distribution in the case I $5.46-0.473 = 4.99$ ($3.9-1.24 = 2.66$) photons, while in case II $96 - 0.342 = 95.7$ ($87 - 1.62 = 85.4$) photons. 
An upper (rough) limit to the DM substructure diffuse $\simeq$ 111 and 129 GeV photon component can be derived by considering it approximately isotropic 
and then counting the $\simeq$ 111 and 129 GeV photons laying above $\mid b \mid \geq 60^{\circ}$ times 7.46 (the ratio of 4$\pi$ to the area of the sky with $\mid b \mid \geq 60^{\circ}$).
There are 40 111$\pm$5 GeV and 30 129$\pm$6 GeV photons above $\mid b \mid \geq 60^{\circ}$, i.e. an upper estimate of the $\simeq$ 111 and 129 GeV 
photons in the isotropic component is 522 photons; thus significantly larger than the 1.0, 4.99 (2.66), or 95.7 (85.4) 
DM substructure diffuse component photons predictions from VLII and unbiased (anti-biased) case I and II.
In section~\ref{sec:LineEmissions} we have a more model-dependent estimate of the isotropic 111 and 129 GeV $\gamma$-rays component, which though decreases the isotropic component photons down to $\simeq$190-230. 

While for the anti-biased distribution we find more events from fewer sources than the unbiased distribution, since the subhalo concentration has a sharper dependence on Galactocentric radius with higher luminous subhalos closer to the GC, here the trend is reversed when summing over the whole population of dim sources.

\begin{table}[t]
\begin{tabular}{|c|c||c|c|}
\hline
Index "$a$"	&	m$_{\textrm{cut}}$ (${\rm M}_{\odot}$)	&	unbiased		& anti-biased \\
\hline \hline
	2.0	&	1.0$\times10^{-6}$			&	96			& 87 \\
\hline
	2.0	&	1.0					&	20.8			& 20.4 \\
\hline
	1.9	&	1.0$\times10^{-6}$			&	16.3			& 10.2 \\
\hline
	1.9	&	1.0$\times10^{3}$			&	5.46			& 3.90 \\
\hline
	1.9	&	2.0$\times10^{4}$			&	4.02			& 2.99 \\
\hline 
\end{tabular}
\caption{Number of 111 and 129 GeV lines photons contributing to the DM subhalo diffuse lines component, for various choices of subhalo distributions, using cross section which fits GC (using cross section which fits whole sky instead, will 
scale all values by a factor of $\simeq3/9.8=0.3$).
We show results for different subhalo mass function spectral index $a$ and lower mass cut-off 
$m_{cut}$, and for unbiased and anti-biased distributions.}
\label{tab:SubhaloDistribution}
\end{table}

\section{Diffuse $\gamma$-ray lines emission from dark matter annihilation}
\label{sec:LineEmissions}

\begin{figure} 
\begin{centering}
\includegraphics[width=3.10in,angle=0]{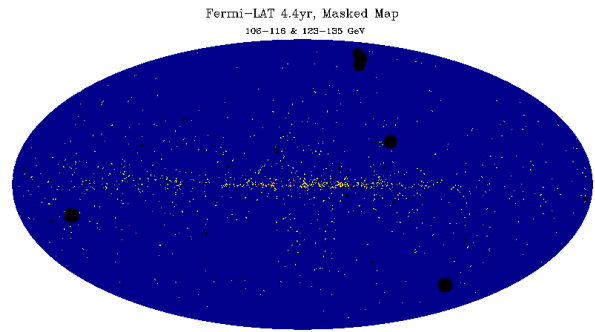} 
\end{centering}
\caption{$\gamma$-ray events (ULTRACLEAN class) with energy of 111$\pm$5 and 129$\pm$6 GeV after 4 yrs 
of collection by \textit{Fermi}-LAT.  We mask out the 16 point sources of  \cite{Su:2012zg} and the 6 extended
sources of \cite{Hektor:2012kc} (see text for more details).
We present $\gamma$-ray events in Mollweide projection using HEALPix \cite{Gorski:2004by}.}
\label{fig:LinesMap}
\end{figure}

As discussed in~\ref{sec:Assumptions} the $\gamma$-ray lines centered at 111 and 129 GeV that are observed in the sky originate 
from combination of sources.
In this part we mask out the contribution from the 16 point sources detected at \cite{Su:2012zg} each with one photon 
at energies between 100-140 GeV. We also mask out the extended sources (galaxy clusters) where 
a similar excess of 100-140 GeV $\gamma$-rays has been observed \cite{Hektor:2012kc}. We use a 
mask of $0.5^{\circ}$ in radius for each of the 16 point sources of \cite{Su:2012zg} and a 
$4^{\circ}$ radius mask for the targets of \cite{Hektor:2012kc}.
The remaining contribution to the 111 and the 129 GeV lines may come from the diffuse 
$\gamma$-rays emission due to DM annihilations in the main halo and its dark disk, from
background $\gamma$-rays produced in the Milky Way, from DM annihilation at small 
scale substructures in the Milky Way (that we have not yet identified as point sources) 
and from other isotropically distributed extragalactic astrophysical sources. 
DM annihilation in extragalactic structures and CR contamination will give an additional 
isotropic component.
In Fig.~\ref{fig:LinesMap} we show the 4$\pi$ sky after implementing our mask on the 16 point
sources and the 6 extended ones.

The DM annihilation rate in any part of the Galaxy is given by:
\begin{eqnarray}
\label{eq:annih_rate}
        \Gamma_{ann}&=&\frac{1}{2 m_{\chi}^{2}}\langle\sigma_{ann}\mid v\mid \rangle \\
&\times&\left(\rho_{sph}^{2} + \rho_{DD}^{2} + 2 \rho_{sph}\cdot \rho_{DD} + \rho_{sub}^{2}\right), \nonumber
\end{eqnarray}
with $\langle\sigma_{ann}\vert v\vert\rangle$ the annihilation cross-section taken to be 
the same for both the DM particles in the dark disk, the spherical halo and the substructures. For the 
case of Sommerfeld enhancement these cross-sections are in general different because of 
the dependence of the annihilation cross-section to the velocity dispersion of dark matter
\cite{ArkaniHamed:2008qn, Lattanzi:2008qa, Hisano:2003ec, Hisano:2004ds} and the fact that 
for the DM particles in the DD the dispersion is suppressed by a factor of 5-6 compared 
to that in the spherical halo components \cite{Read:2008fh}; with subhalos having even lower 
velocity dispersions.
Thus for Sommerfeld enhanced models the dark disk contribution to CRs and $\gamma$-rays 
can be much more significant \cite{Cholis:2010px} (see also \cite{Slatyer:2011kg} for a discussion on the impact of subhalos).

In fitting to the 2 $\gamma$-ray line full sky data we probe the prompt 
$\gamma$-ray DM annihilation component of the spectrum which is directly related to 
the annihilation rate in eq.~\ref{eq:annih_rate}.

\begin{figure*} 
\begin{centering}
\includegraphics[width=3.10in,angle=0]{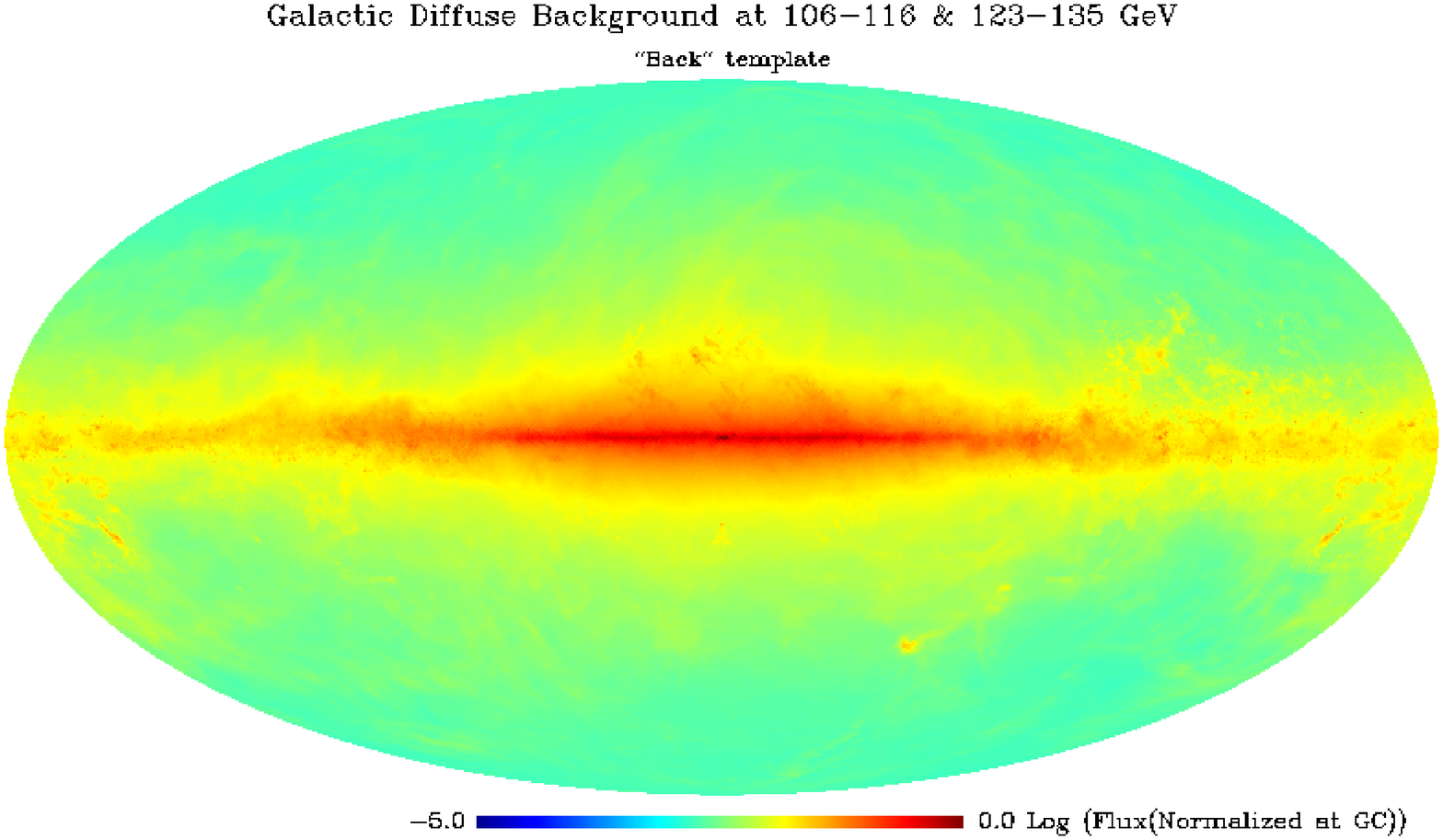}
\includegraphics[width=3.10in,angle=0]{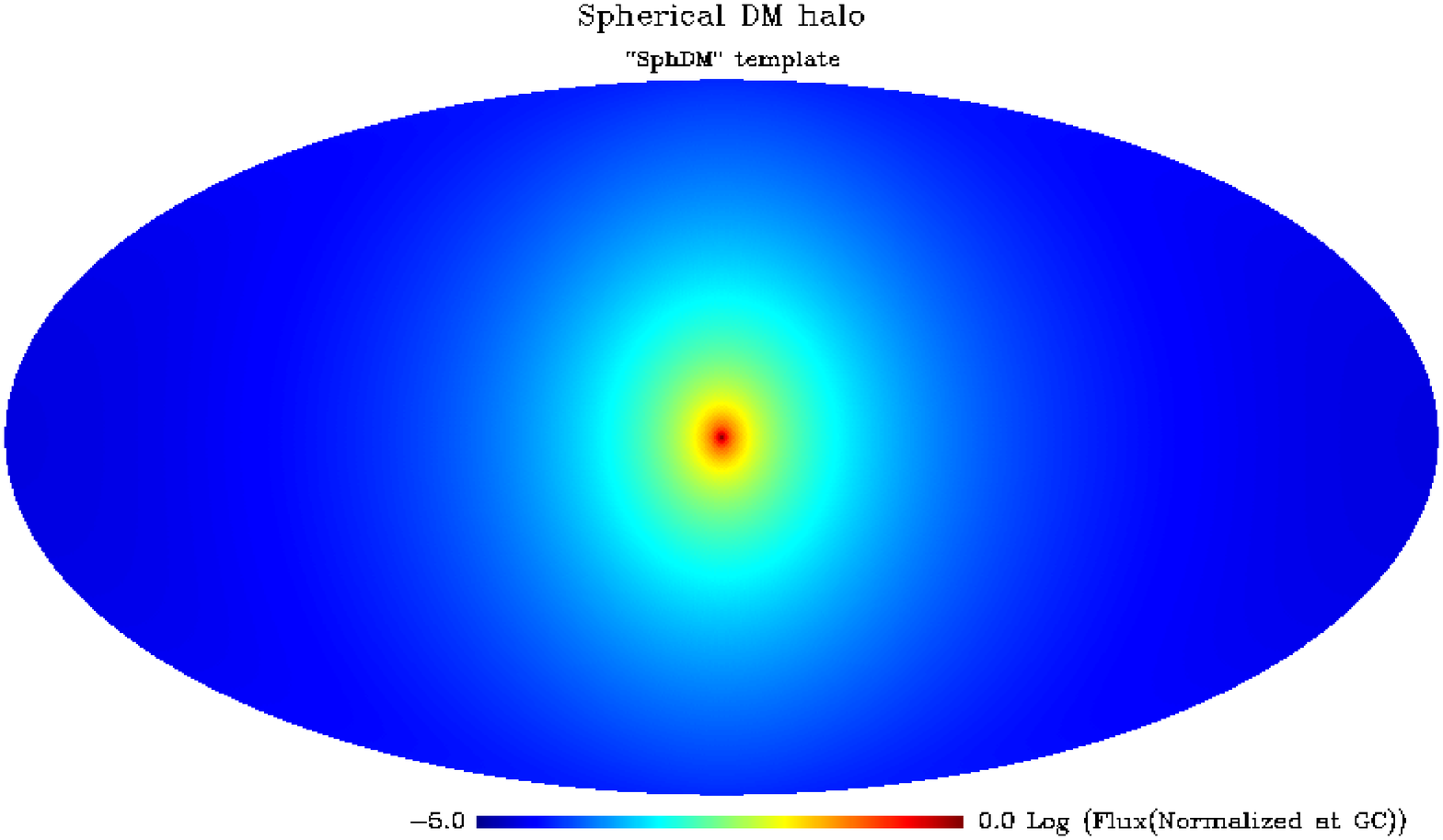} \\
\includegraphics[width=3.10in,angle=0]{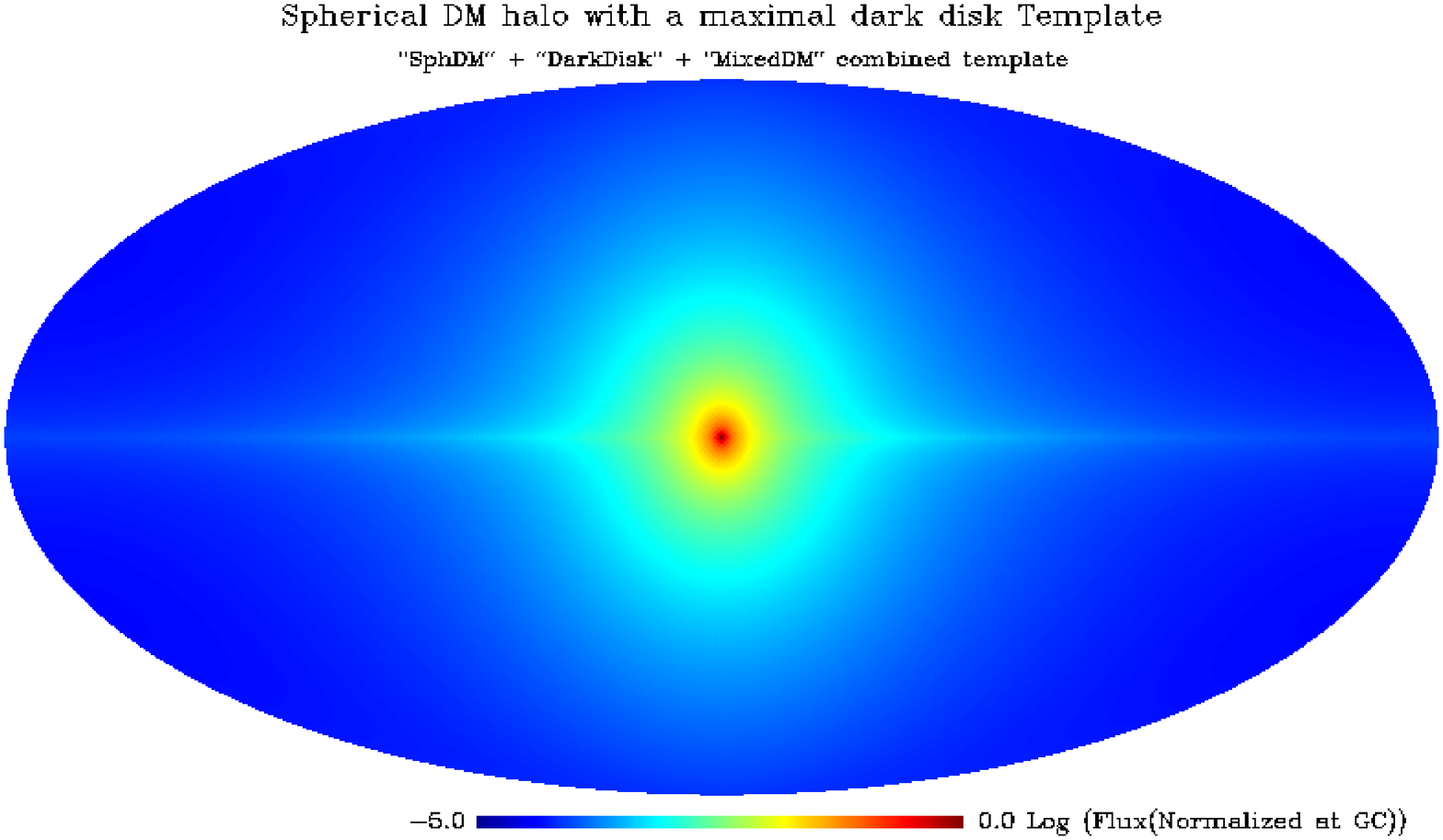}
\includegraphics[width=3.10in,angle=0]{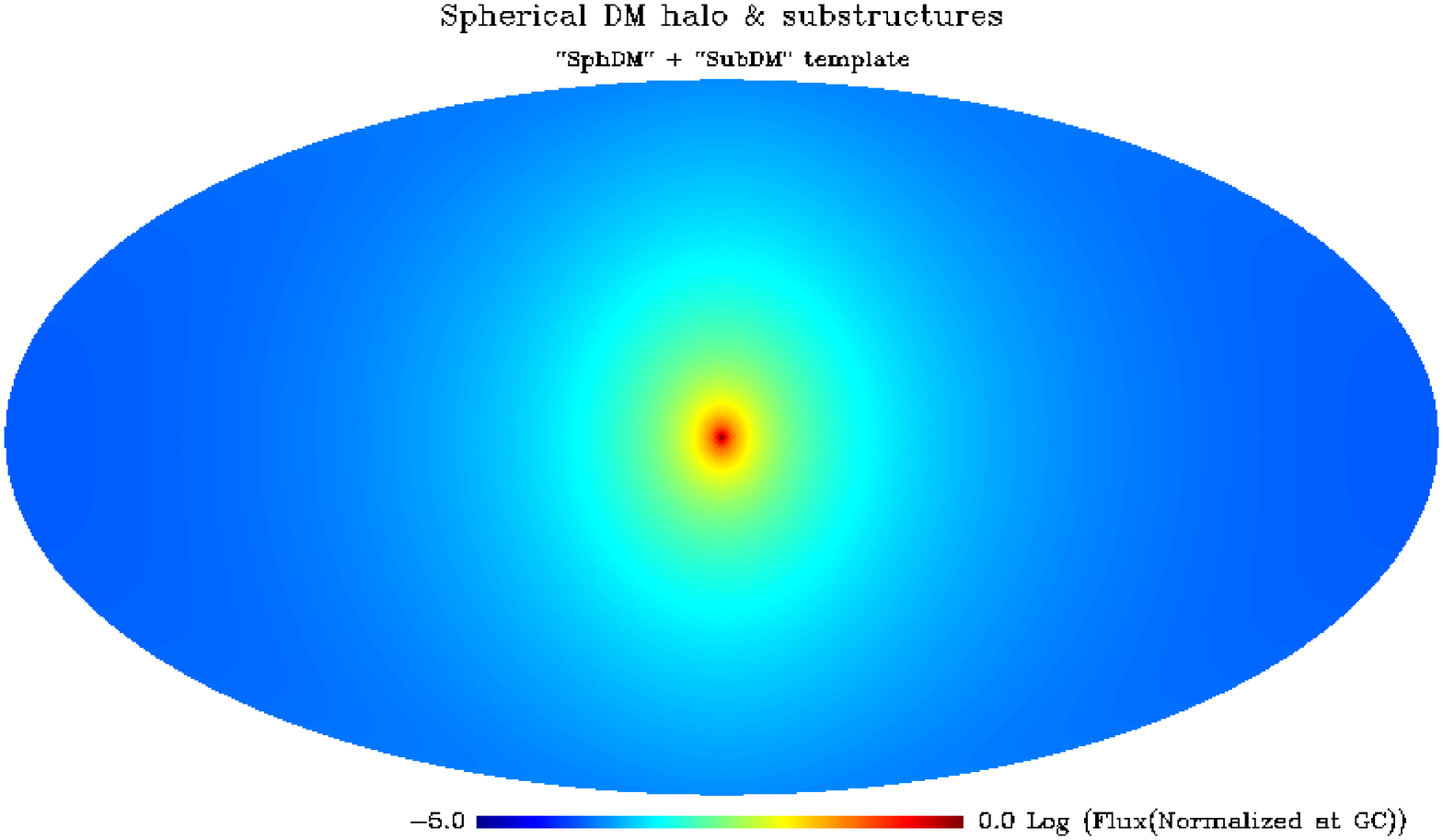}
\end{centering}
\caption{The $\gamma$-ray templates as can be used in eq.~\ref{eq:Model}. \textit{Top left}: the Galactic diffuse background
template including the $\pi^{0}$, inverse Compton scattering and bremsstrahlung components 
at energies of 111$\pm$5 and 129$\pm$6 GeV. We show the "Back A" model. \textit{Top right}: the diffuse DM spherical halo component assuming 
an Einasto with $\delta = 0.13$ profile. \textit{Bottom left}: diffuse emission from the combined spherical and dark disk  
DM distributions, assuming a maximal dark disk ($\alpha =1$), which results in showing the "SphDM"+ "DarkDisk" + "MixedDM" 
combined template. We use Einasto profile with $\delta = 0.13$ for the spherical 
and $z_{1/2} = 0.5$ kpc for the scale hight of the dark disk. \textit{Bottom right}: the DM diffuse subhalo template "SubDM" for 
unbiased distribution with $m_{cut} = 10^{-6} M_{\odot}$ together with the spherical DM halo "SphDM" template (Einasto with $\delta = 0.13$) .
We use Mollweide projection. To demonstrate the different morphologies, each template is normalized to 1 at the GC. Dark blue (dark grey) color 
refers to a flux suppressed by a factor of $10^{-5}$ compared to the GC in each template.  
In eq.~\ref{eq:Model} we use the calculated flux values from the DRAGON package.}
\label{fig:Templates}
\end{figure*}

We use the masked full sky data with energies $111 \pm 5$ GeV and $129 \pm 6$ GeV.
We perform a maximum likelihood fit calculating the log-likelihood based on
\cite{Dobler:2009xz}:
\begin{equation}
  \ln {\mathcal L} = \sum_i k_i\ln\mu_i - \mu_i - \ln(k_i!),
\label{eq:loglikelihood}
\end{equation}
where $\mu_i$ is the model of linear combination of
templates at pixel $i$, and $k$ is the map of observed counts which is just the single 
$111 \pm 5$ GeV and $129 \pm 6$ GeV $\gamma$-ray \textit{Fermi} masked map.
Our \textit{diffuse} $\gamma$-ray model is composed of 6 templates with 4 free parameters:
\begin{eqnarray}
\label{eq:Model}
  \mu_i &=& N\cdot \textrm{Back}_{i} + A \cdot [(2-\alpha)^{2} \cdot \textrm{SphDM}_{i} \\
   &+& \alpha^{2} \cdot \textrm{DarkDisk}_{i} + \alpha(2-\alpha) \cdot \textrm{MixedDM}_{i} \nonumber \\
   &+& \textrm{SubDM}_{i}] + B \cdot \textrm{Iso}_{i}. \nonumber
\end{eqnarray}
The Back$_{i}$ template comes from our DRAGON run and is kept fixed modulo a normalization
$N$ for a specific set
of assumptions on the ISM gas and ISRF energy densities, the SphDM$_{i}$ refers to the term in eq.~\ref{eq:annih_rate}
that is proportional to $\rho_{sph}^{2}$. The DarkDisk$_{i}$ template refers to the term 
in eq.~\ref{eq:annih_rate} proportional to $\rho_{DD}^{2}$ and the MixedDM$_{i}$ template 
to the $2 \rho_{sph}\cdot \rho_{DD}$ term. 
The contribution of dim Galactic DM subhalos to the diffuse $\gamma$-rays is included in the "SubDM" (related to the $\rho_{sub}^2$ term in eq.~\ref{eq:annih_rate}),
where as described earlier we have masked out the brightest possibly detected members.  
$\alpha$ refers to the ratio of local DM density 
of the DD over the spherical halo given in eq.~\ref{eq:init_contraints2}. 
The Iso$_{i}$ template includes the contribution of DM extragalactic annihilations, the extragalactic background from other 
sources and the possible CR contamination. We also multiply the model map by the total FRONT+BACK-converted ULTRACLEAN photons 
exposure map and multiply by the mask.
In Figure~\ref{fig:Templates} for specific choices we show 4 different templates. On top left a specific model ("Back A") for the Galactic diffuse 
background is shown at the energies of interest. On top right we plot the "SphDM" template for an Einasto DM spherical halo profile 
with $\delta = 0.13$, and on bottom left we show the combined DM spherical and dark disk for $\alpha =1$ ("SphDM"+ "DarkDisk" + "MixedDM"). 
The impact of adding the DM subhalos contribution ("SubDM") is given in the bottom right. 
We use HEALPix \cite{Gorski:2004by} with Nside  = 128 which represents closer the angular 
resolution of \textit{Fermi} LAT at these energies \footnote{The angular resolution of the \textit{Fermi} 
LAT instrument is different at the energies that we study by about a factor of 2 between the front 
and the back-converted converted photons. With Nside=128 our pixels include entirely the front-converted 
photons and are slightly smaller than the $68\%$ containment angle for back-converted events. Choosing 
Nside=256 would have resulted in photon events being spread in multiple pixels thus making them correlated;
while Nside$\leq 64$, would be underestimating the ability of the instrument to resolve $\gamma$-ray structures
at these energies.  We have checked that our 
best fit results are consistent within 2$\%$ for Nside of 128, 64, 32, 16 and 8.}.

\begin{center}
\begin{table*}[ht]
\begin{tabular}{|c||c|c||c|c|c||c|}
\hline
DM profiles / Backgrounds & $\sigma v$ & $F_{iso}$  & Back ph.& DM ph.& Iso ph. & TS\\
\hline \hline
Ein. ($\delta = 0.13$)  / Back A&  1.5 (4.5) & 5.73 & 1146 & 40 (121) & 214 & 9.1 \\
\hline
Ein. ($\delta = 0.17$)  / Back A&  2.2 (7.1) & 5.55 & 1146 & 43 (138) & 207 & 6.1 \\
\hline
Ein. ($\delta = 0.22$)  / Back A&  2.7 (8.5) & 5.38 & 1157 & 41 (127) & 201 & 2.8 \\
\hline
Ein.  ($\delta = 0.13$) / Back B& 1.6 (4.8) & 5.87 & 1134 & 44 (129) & 219 & 11.9 \\
\hline
Ein. ($\delta = 0.13$) / Back C& 1.5 (4.6) & 5.81 & 1144 & 39 (124) & 217 & 9.2 \\
\hline
Ein. ($\delta = 0.13$) / Back D& 1.3 (4.3) & 6.05 & 1137 & 36 (115) & 226 & 7.8 \\
\hline 
\end{tabular}
\caption{The values of relevant parameters for various assumptions on the Galactic diffuse background and the distribution of DM spherical halo ignoring the contribution of a DD ($\alpha=0$) and subhalos. 
Second column gives the best fit annihilation cross-section assuming equal annihilation cross-sections to the 
two lines; $\langle\sigma v\rangle _{\gamma Z}$ = $\langle\sigma v\rangle _{\gamma \gamma}$ $\equiv$ 
$\sigma v$ in units of $\times 10^{-28}$ cm$^{3}$ s$^{-1}$.
 $F_{iso}$ is the calculated isotropic flux at $111 \pm 5$ GeV and $129 \pm 6$ GeV  in
units of $\times 10^{-12}$ GeV$^{-1}$ cm$^{-2}$ s$^{-1}$ sr$^{-1}$.
Columns  4-6 refer to the $111 \pm 5$ GeV and $129 \pm 6$ GeV 
photons in the Background, DM and Isotropic diffuse components as predicted by the fitted values of $N$, $A$ and $B$ of eq.~\ref{eq:Model}. 
Last column gives the TS for detection of a DM signal. Values in parentheses refer to 3$\sigma$ upper limits on the DM annihilation cross-section.}
\label{tab:LikelihoodFits}
\end{table*}
\end{center}

\begin{center}
\begin{table*}[ht]
\begin{tabular}{|c||c|c||c|c|c||c|}
\hline
DM profiles / Backgrounds & $\sigma v$ & $F_{iso}$  & Back ph.& DM ph.& Iso ph. & TS \\
\hline \hline
Ein. ($\delta = 0.13$); DD $z_{1/2} = 0.5$ / Back A& 2.8 & 5.72 & 1143 & 43 & 213 & 8.7 \\
\hline
Ein. ($\delta = 0.13$); DD $z_{1/2} = 1.0$ / Back A& 2.6  & 5.69 & 1144 & 42 & 212 & 8.0 \\
\hline
Ein. ($\delta = 0.13$); DD $z_{1/2} = 1.5$ / Back A& 2.5  & 5.64 & 1146 & 43 & 210 & 7.7 \\
\hline
Ein. ($\delta = 0.13$); DD $z_{1/2} = 3.0$ / Back A& 2.4  & 5.60 & 1145 & 45 & 209 & 7.6 \\
\hline
Ein. ($\delta = 0.17$); DD $z_{1/2} = 0.5$ / Back A& 4.2  & 5.56 & 1143 & 49 & 208 & 5.6 \\
\hline
Ein. ($\delta = 0.22$); DD $z_{1/2} = 0.5$ / Back A& 4.7  & 5.40 & 1154 & 43 & 201 & 2.4 \\
\hline
Ein. ($\delta = 0.13$); DD $z_{1/2} = 0.5$ / Back B& 3.1  & 5.91 & 1130 & 48 & 221 & 11.5 \\
\hline
Ein. ($\delta = 0.13$); DD $z_{1/2} = 0.5$ / Back C& 2.8  & 5.79 & 1141 & 43 & 216 & 9.0 \\
\hline
Ein. ($\delta = 0.13$); DD $z_{1/2} = 0.5$ / Back D& 2.5  & 6.04 & 1135 & 38 & 225 & 7.5 \\
\hline 
\end{tabular}
\caption{The values of relevant parameters for various assumptions on the Galactic diffuse background, the distribution of DM spherical halo and the DD assuming the maximal DD contribution ($\alpha=1$) and ignoring the contribution of subhalos. 
Second column is as in Table~\ref{tab:LikelihoodFits}, 
$F_{iso}$ is in units of $\times 10^{-12}$ GeV$^{-1}$ cm$^{-2}$ s$^{-1}$ sr$^{-1}$ and $\sigma v$ is 
in units of $\times 10^{-28}$ cm$^{3}$ s$^{-1}$.
Columns  4-6 refer to the $111 \pm 5$ GeV and $129 \pm 6$ GeV 
photons in Back, DM and Iso components predicted by the fitted values of $N$, $A$ and $B$. 
Last column gives the TS for detection of DM signal.}
\label{tab:LikelihoodFitsDD}
\end{table*}
\end{center}

We also calculate the significance of a DM contribution from the diffuse analysis by the test statistic, where 
\begin{equation}
  TS \equiv -2 ln \frac{\mathcal L_{null}}{\mathcal L_{best fit}}.
 \label{eq:TestStat} 
\end{equation}
$\mathcal L_{best fit}$ allows for the DM to contribute, while in $\mathcal L_{null}$ we set the DM 
diffuse component to zero and refit the other two diffuse components.
Our results are shown in Tables~\ref{tab:LikelihoodFits}-\ref{tab:LikelihoodFitsSubHalos}.

Ignoring first both the contribution of a dark disk and the DM subhalos, we find that the more cuspy DM profiles for the 
main/spherical halo that lead to less DM contribution to the diffuse 
$\gamma$-ray spectrum away from the Galactic center, provide a larger positive fit to the 4$\pi$ sky (see Table~\ref{tab:LikelihoodFits}). 
Yet the significance of that is not very large (TS = 9.1/9.2 and 11.9 in the best cases).  
An even more cored (flat) Burkert DM profile for the main 
halo in the inner kpcs does not give a better fit to the 111 and 129 GeV lines distribution in agreement with the 
findings of \cite{Su:2012ft} and \cite{Weniger:2012tx} performed in subsections of the sky. 

Using the model of $\delta = 0.13$ for the Einasto DM density profile that provides the best fit, 
we also test different diffuse background models ("Back A"-"Back D") to account for uncertainties in the interstellar medium gas 
distribution and interstellar radiation field distribution. We find that in all cases a DM component  is preferred by the fit at $\simeq$2-3 
$\sigma$ significance for the cuspier DM models (1-sided since in our fits we allow for the DM component to be even negative), and accounting 
for about 35-45 photons (cross-sections of $\langle \sigma v \rangle = 1.3 -2.7 \times 10^{-28}$ cm$^{3}$s$^{-1}$) 
\footnote{Since we use the entire 4$\pi$ sky and have selected in advance the energy range of the $\gamma$-ray events based on the 
works of \cite{Weniger:2012tx, Su:2012ft, Tempel:2012ey, Rajaraman:2012db, Buchmuller:2012rc, Boyarsky:2012ca} there is no look 
elsewhere effect accounted for.}. As with "Back A" the cuspier DM profiles are preferred for the tested backgrounds.
Our "Back A" is the same as the reference model described in \cite{MaryamEtAl},
which was shown to provide a good agreement to the 4$\pi$ sky and in energies between 1 and 200 GeV and has also been cross-checked to local CR measurements.  
Model "Back B" assumes a different distribution for the molecular hydrogen gas component that is dominant at lower latitudes and toward the GC where 
many of the 111 and 129 GeV photon excess have been claimed (see \cite{MaryamEtAl} for more details).  
"Back C" and "Back D" Galactic diffuse models assume, respectively, an enhanced ISRF energy density distribution toward the disk 
and a minimal metallicity gradient \cite{Porter:2008ve}.
The latter assumption affects both the morphology and the spectrum of the ISRF and as a result the 
$\gamma$-rays produced via Compton up-scattering of these photons by high energy CR electrons. These background models are discussed in further detail in \cite{MaryamEtAl}. They have not been constructed to study just the Galactic $\gamma$-ray background at $\sim$110-130 GeV, but instead the general uncertainties in the Galactic diffuse $\gamma$-rays between 1-200 GeV, in the full sky and in subsections of it.    

Since in our fits we also allow for an isotropic component we can calculate the isotropic flux at these energy ranges,
taking into account also the Galactic diffuse background uncertainties. We find that $\simeq 210-230$ photons can be 
accounted by that component. This gives an isotropic flux of 5.6-6.1 $\times 10^{-12}$ GeV$^{-1}$ cm$^{-2}$ s$^{-1}$ 
sr$^{-1}$ which is in agreement with the extragalactic isotropic $\gamma$-ray flux of \cite{Abdo:2010nz} described by 
$dN/dE \propto E^{-2.41 \pm 0.05}$ measured between 200 MeV and 100 GeV. The  $\simeq 210-230$ photons of 
the isotropic component can also be used to set approximate limits on the contribution from  DM halos  at extragalactic 
distances.

We find that adding a dark disk component does not significantly change the fit to the data from the case of having only a spherical DM halo contribution.
That is for any choice of diffuse background or spherical DM halo shown Table~\ref{tab:LikelihoodFitsDD}. Thus a significant DD contribution is not preferred by the current data; while yet it can not be excluded either. We give the likelihood fits for the maximal DD contribution.  
\begin{center}
\begin{table*}[ht]
\begin{tabular}{|c||c|c|c||c|c|c||c|}
\hline
DM profiles & $\sigma v$ & $\alpha$ & $F_{iso}$  & Back ph.& DM ph.& Iso ph. & TS\\
\hline \hline
Ein. ($\delta = 0.13$); Unbiased Sub. Dist. ($m_{cut}=10^{-6}M_{\odot}$) & 1.4 & 0.0 & 5.34 & 1146 & 53(16) & 199 & 8.6 \\
\hline
Ein. ($\delta = 0.13$); Unbiased Sub. Dist. ($m_{cut}=10^{+6}M_{\odot}$) & 1.5 & 0.0 & 5.70 & 1146 & 40(0.36) & 213 & 9.1 \\
\hline
Ein. ($\delta = 0.13$); Anti-Biased Sub. Dist. ($m_{cut}=10^{-6}M_{\odot}$) & 1.4 & 0.0 & 5.39 & 1146 & 52(14) & 201 & 8.8 \\
\hline
Ein. ($\delta = 0.13$); Anti-Biased Sub. Dist. ($m_{cut}=10^{+6}M_{\odot}$) & 1.5 & 0.0 & 5.70 & 1146 & 40(0.41) & 213 & 9.1 \\
\hline
Ein. ($\delta = 0.17$); Unbiased Sub. Dist. ($m_{cut}=10^{-6}M_{\odot}$) & 2.0 & 0.0 & 5.07 & 1149 & 61(23) & 189 & 5.2 \\
\hline
Ein. ($\delta = 0.17$); Unbiased Sub. Dist. ($m_{cut}=10^{+6}M_{\odot}$) & 2.4 & 0.0 & 5.54 & 1146 & 46(0.59) & 207 & 6.1 \\
\hline
Ein. ($\delta = 0.17$); Anti-Biased Sub. Dist. ($m_{cut}=10^{-6}M_{\odot}$) & 2.1 & 0.0 & 5.10 & 1149 & 60(21) & 190 & 5.5 \\
\hline
Ein. ($\delta = 0.17$); Anti-Biased Sub. Dist. ($m_{cut}=10^{+6}M_{\odot}$) & 2.4 & 0.0 & 5.54 & 1146 & 47(0.66) & 207 & 6.1 \\
\hline
Ein. ($\delta = 0.13$); DD $z_{1/2} = 0.5$; Unbiased Sub. Dist. ($m_{cut}=10^{-6}M_{\odot}$) & 2.4 & 1 & 5.04 & 1144 & 64(28) & 188 & 7.8 \\
\hline
Ein. ($\delta = 0.13$); DD $z_{1/2} = 0.5$; Unbiased Sub. Dist. ($m_{cut}=10^{+6}M_{\odot}$) & 2.8 & 1 & 5.769 & 1143 & 43(0.69) & 212 & 8.7 \\
\hline
Ein. ($\delta = 0.13$); DD $z_{1/2} = 0.5$; Anti-Biased Sub. Dist. ($m_{cut}=10^{-6}M_{\odot}$) & 2.6 & 1 & 5.13 & 1143 & 64(25) & 192 & 8.2 \\
\hline
Ein. ($\delta = 0.13$); DD $z_{1/2} = 0.5$; Anti-Biased Sub. Dist. ($m_{cut}=10^{+6}M_{\odot}$) & 2.8 & 1 & 5.69 & 1143 & 43(0.77) & 212 & 8.7 \\
\hline 
\end{tabular}
\caption{The values of relevant parameters for various assumptions on the distribution of DM spherical halo, the DD and dim Galactic DM subhalos with different choices of the minimal subhalo mass. 
Units for $\sigma v$ and $F_{iso}$ are as in Tables~\ref{tab:LikelihoodFits} and \ref{tab:LikelihoodFitsDD}. 
In parentheses we give the photon number associated to the subhalo diffuse component only.  The subhalo mass spectral index is assumed to be -2.
In all cases we use our reference "BackA" as Galactic diffuse background.} 
\label{tab:LikelihoodFitsSubHalos}
\end{table*}
\end{center}
Our fits give a relatively flat behavior in the parameter $\alpha$ with a slight preference for  $\alpha \longrightarrow 0$.
More specifically for a DD with $z_{1/2} \geq 1.5$ kpc 
(thick DD) with $\alpha \longrightarrow 1$ we get a lower test statistic by about $\Delta TS \simeq 2$ compared to the cases of  having only the  spherical DM halos contributing, i.e. the data show a slight preference for having only a spherical DM halo at these energies. 
That is for any of our background assumptions. Yet thin DD  ($z_{1/2} \simeq 0.5$ kpc) give about the same TS as the only spherical DM halo contribution for any of our background models. 
Having tested also the 4 yr sample, that gave a slight preference for $\alpha =1$ with a  $z_{1/2} \simeq 0.5$ kpc DD, we conclude that the current  small number of photons at the two energy lines does not provide us with enough detail to discriminate a thin DD component  from the disk-like distribution of the Galactic background $\gamma$-rays.  With more $\gamma$-ray data we expect that we will be able to further disentangle the different $\gamma$-ray morphologies.

Even for the case where we have a maximal DD contribution, i.e. when $\alpha =1$ most of the DM photons are 
due to the spherical and the mixed terms of eq.~\ref{eq:annih_rate}. In those cases,  the equivalent DM annihilation cross-section that accounts for the DM template normalization is maximal. That is the case since the total number of DM photons is about the same while the $\left(\rho_{sph}^{2} + \rho_{DD}^{2} + 2 \rho_{sph}\cdot \rho_{DD}\right)$ term in eq.~\ref{eq:annih_rate} is minimal.  Since the DD term can not be excluded this is a way of allowing for higher annihilation cross-sections, by up to a factor of $\simeq 2$.

In Table~\ref{tab:LikelihoodFitsSubHalos} we also study the impact of the diffuse DM subhalo template "SubDM" of eq.~\ref{eq:Model}
with and without a DD component. To bracket the uncertainties we used both the unbiased and the anti-biased Galactic subhalo distributions
(see Appendix). We also tested the subhalo mass extrapolations down to $10^{+6} M_{\odot}$ and $10^{-6} M_{\odot}$.
For the cases where the extrapolation is down to $10^{+6} M_{\odot}$ the DM subhalo diffuse component is subdominant contributing
only a few line photons. On the contrary, for extrapolations down to $10^{-6} M_{\odot}$ that component can contribute up to $\sim 1/3$ ($\sim 1/2$)
of the Galactic DM line photons, with the remaining coming from the spherical main halo (spherical $\&$ DD components).
With the current data the difference in the TS fit between the unbiased and the anti-biased distributions is very small (see Table~\ref{tab:LikelihoodFitsSubHalos}).
Yet, even adding a strong subhalo term/template, our method can discriminate between different assumptions for the cuspiness of the spherical halo "SphDM".
In agreement with all our previously discussed tests, a preference toward cuspier halos is found. In Table~\ref{tab:LikelihoodFitsSubHalos} we compare 
between $\delta = 0.13$ and $\delta = 0.17$ Einasto profiles that differ only in the inner few degrees from the GC.

Having tested the DM case both with and without a dark disk and including/excluding the contribution of Galactic bound dim DM subhalos we 
have consistently found a thermally averaged cross-section $\langle \sigma v \rangle$$\equiv$$\langle \sigma v \rangle_{\gamma \gamma}$
=$\langle \sigma v \rangle_{\gamma Z}$ for the two lines that is in the range of 1.5-4.5 $\times 10^{-28}$ cm$^{3}$s$^{-1}$. These values are a factor 
of 9-3 smaller than the suggested values from analyses coming when concentrating only toward the GC \cite{Weniger:2012tx}. Fitting the entire 
4$\pi$ sky can dilute the DM signal from the GC and thus suggest a smaller annihilation cross-section. Yet, we note that a strong annihilation 
annihilation cross-section to the lines should be seen at the diffuse spectrum at high latitudes when including the contribution from the dim subhalos 
with masses down to the free  streaming  scale. Our fits do not suggest such a case. Taking the subhalos to have masses down to $10^{-6}$$M_{\odot}$, 
we get from our fits shown in Table~\ref{tab:LikelihoodFitsSubHalos} $\simeq$ 15-20 photons (30 including a DD in the fit) from annihilations just in the 
dim subhalos. That number of photons from the diffuse subhalo component can be explained by a thermally averaged cross-section of 1.5-2.5
$\times 10^{-28}$  cm$^{3}$s$^{-1}$. Given that the velocity dispersion in bound substructures is smaller than in the GC or locally that may be expected. 
Alternatively explained, thermally averaged cross-section of $\langle \sigma v \rangle = 1.0 \times 10^{-27}$ cm$^{3}$s$^{-1}$ down to the smallest subhalos
 would give $\sim$ 100 photons on the 4$\pi$ sky just on the two lines and just from the dim Galactic subhalos. Of these $\simeq 25$ photons would be at 
 $\mid b\mid \ge 45^{\circ}$ with the isotropic component predicting $\simeq 70$ photons in the two energy ranges ($111\pm 5$, $129\pm 6$ GeV). Thus the 
 lines would have to be observed at high latitudes as well. If not found at these latitudes, then either the $\langle \sigma v \rangle$ is smaller in the subhalos, 
 in general, or subhalos are not formed down to masses of $\sim 10^{-6}$$M_{\odot}$. We note that in the above numbers we have not included the possible 
 and more model dependent contribution form extragalactic DM annihilations.  

\section{Discussion and Conclusions}
\label{sec:Conclusions}
Recently, \cite{Su:2012zg} has found indications for the 111 and the 129 GeV lines in the 2 yr \textit{Fermi} unassociated point sources 
catalogue that would be indicative for DM annihilation in substructures.
We compare the compatibility of the findings of \cite{Su:2012zg} with the results from VLII cosmological simulations and  extrapolations of it. In that process we 
have assumed the same annihilation cross-section to the Galactic center lines signal \cite{Weniger:2012tx, Su:2012ft, Tempel:2012ey, 
Rajaraman:2012db, Buchmuller:2012rc, Boyarsky:2012ca} given that the point sources lines signal and the Galactic center line(s) signal have been observed 
at the same $\gamma$-ray energies. 
We find that just considering VLII assumptions we do not get enough line photons from the brightest Galactic DM subhalos to claim agreement 
with the signal seen by \cite{Su:2012zg}. 
Yet completions of the VLII simulation results do give a number of line photons from the brightest subhalos that in the most optimistic cases is in
good agreement with (but still below) the lines events number found by \cite{Su:2012zg} (9 events in their ULTRACLEAN sample) (see discussion in section~\ref{sec:VLII}). 
These most optimistic extrapolations to very small substructures ($10^{-6} M_{\odot}$) predict that many diffuse line photons that are in some tension with the $\gamma$-ray flux at high latitudes, but still can not be excluded in the most conservative manner. The same applies when we compare their predictions to the isotropic $\gamma$-ray flux which for the energies of the lines 
we have calculated to be centered at $5.6 \pm 0.3$ $\times 10^{-12}$ GeV$^{-1}$ cm$^{-2}$ s$^{-1}$ sr$^{-1}$, with 4$\pi$ sky fit values being in the range of $5.0-6.0$ 
$\times 10^{-12}$ GeV$^{-1}$ cm$^{-2}$ s$^{-1}$ sr$^{-1}$ (accounting for systematic/model uncertainties in the evaluation of the isotropic component).

Most of the 111 and the 129 GeV line photons on the $\gamma$-ray sky are not from DM annihilations but rather from the Galactic diffuse and the isotropic diffuse background (shown in Fig.~\ref{fig:LinesMap}). By doing a template fit to the 4$\pi$ sky, we test how robust the DM signal hypothesis is for different physical assumptions on the Galactic diffuse $\gamma$-ray background flux and on the DM halo profile (see discussion on section~\ref{sec:LineEmissions}). We find a positive fit of having a DM spherical halo component.   
Our DM annihilation component hypothesis is preferred to the case without a DM component at a test statistic significance of up to 12; with the values depending on the exact DM halo assumptions. More concentrated DM profiles give a larger significance to a DM signal. 
We also find that our results on the positive signal of DM annihilation are weakly dependent on Galactic diffuse background uncertainties
related to either the uncertainties in the distribution of the interstellar medium gas or to the energy density in the interstellar radiation field.

Extending our set of tests on the DM distribution in the Galaxy, we include a dark disk component that could explain the non-isotropic distribution on the sky of the  
point sources of  \cite{Su:2012zg}. Our fits  can not strongly favor or disfavor a significant disk-like DM component, even though there is a small preference toward 
thinner dark disks. 
We also study the contribution of the dimmest DM subhalos in the Milky Way to the diffuse gamma-ray sky at energies around 111 and 129 GeV and find a  preference toward an anti-biased distribution of the subhalos within the main spherical DM halo. Yet our analysis is somewhat constrained in its power by the small number of 
photons on the sky at the energies of the 2 lines. As more statistics are being accumulated, a better understanding of the morphology of the $\gamma$-ray sky will be achieved 
allowing for  such a template analysis to further disentangle the background $\gamma$-ray sky from any possible  DM component.   

Finally using the various combinations of backgrounds and DM distributions, we find the thermally averaged annihilation cross-section to be smaller than what has 
been originally suggested, 
with values ranging between $1.5-4.5 \times 10^{-28}$ cm$^{3}$s$^{-1}$. While the full sky fits are not optimal for a DM signal toward the GC, 
they include high latitudes which probe also the contribution from smaller substructures. 
A suppressed flux at  high latitudes to the lines either indicates a smaller overall cross-section or a suppressed contribution from smaller substructures compared to the GC.

\vskip 0.2 in
\section*{Acknowledgments}  
The authors would like to thank Alex Drlica-Wagner, Carmelo Evoli, Andrew Hearin, Dan Hooper, Ran Lu, Meng Su and Gabrijela Zaharijas 
for valuable discussions. PU acknowledges partial support from the European Union FP7
ITN INVISIBLES (Marie Curie Actions, PITN-GA-2011-289442).
In this work the authors have used the publicly available \textit{Fermi}-LAT data and Fermi Tools
archived at \texttt{http://fermi.gsfc.nasa.gov/ssc/}.
We also acknowledge the use HEALPix. 
\vskip 0.05in

\begin{appendix}

\section{DM Signal from Unbiased and Anti-Biased Distributions of Subhalos in the Milky Way}
\label{sec:UnBiasedDistr}

We fit the subhalos in the VLII simulation \cite{Diemand:2008in,VLIIdata} using an NFW profile \cite{Kuhlen:2008aw}:
\begin{equation}
\rho_{sub}(r)=\rho_{s}\left(\frac{r}{r_{s}}\right)^{-1}\left(1+\frac{r}{r_{s}}\right)^{-2}\;.
\end{equation}
We read out the scale radius, $r_s$, and scale density, $\rho_s$, in terms of the peak of circular 
velocity $V_{max}$ and the corresponding radius $r_{Vmax}$ (see, e.g. equation 9-10 in \cite{Kuhlen:2008aw}).
Their luminosities are given by:
\begin{equation}
 L = 4/3\pi r_{s}^{3}\rho_{s}^{2}\left[1-\left(1+r_{t}/r_{s}\right)^{-3}\right]\;,
\end{equation}
where $r_t$ is their tidal radius. By using this formula, combined with eq. \eqref{eq:numgam}, 
we find the results in the first row of Table~\ref{tab:SubhaloContribution}.

In deriving the unbiased distribution of subhalos, first, we calculate the gravitational potential from the host halo, 
cutting the profile at its virial radius and distinguish bound subhalos from the unbound ones. 
There are 13510 bound subhalos, 9372 of them are within the simulated galaxy's virial radius, 
Furthermore, there are only 9 unbound subhalos within the virial radius and all of them are beyond 250 kpc from GC. 
We also calculate each subhalo's pericenter, $r_p$, to obtain the pericenter probability distribution function (PDF) 
as a function of Galactocentric radius, $r_g$, e.g $\mathrm{d}P/\mathrm{d}r_p (r_p,r_g)$.
For \emph{unbound subhalos}, we identify their tidal radius
as their \emph{virial radius}, and their tidal mass as their \emph{virial mass}. 
According to the virial theorem, the average density inside the virial radius
should be the same for all halos. We find that this is indeed the case in VLII simulation: 
the average density inside the virial radius of unbound subhalos is $\Delta_0\rho_{crit}$, where $\Delta_0$ is the
overdensity relative to the critical matter density for spherical collapse for $z=0$ and $\rho_{crit}$ is the critical \emph{matter} density. 
For bound subhalos, the tidal radius is usually much smaller than virial radius because of tidal stripping. 
We define the tidal concentration of a subhalo, $c_t$, as its average density within its tidal radius
divided by its critical matter density; for unbound subhalos, $c_t=\Delta_0$.
To determine the virial concentration PDF for unbound subhalos,
we use the Bullock model \cite{Bullock:1999he} as parametrized in eq. (2) of \cite{Zechlin:2011kk} 
which relates the median virial concentration of a subhalo,
$c_{vir}=r_{vir}/r_{s}$, with its virial mass, $m_{vir}$.
We find that this parametrization can fit \emph{unbound subhalos's} virial concentrations fairly well.
We take the mass PDF of \emph{unbound} subhalos to be a power law, $\mathrm{d}P/\mathrm{d}m_t\propto m_t^{-a} \Theta(m_t-m_{cut})$ 
and vary the spectral index $a=[1.9,2]$. In fitting the mass and concentration PDF, we use the unbound subhalos 
because they are not tidally stripped. Hence, in this model, the Galactocentric radial dependence of mass and concentration
will appear later, after they are tidally stripped by their host.
To find a subhalo's minimum concentration at some Galactocentric radius $r_g$, we apply the Roche criteria:  
for a subhalo in circular orbit, the subhalo's self-gravity at $r_t$ should be equal to 
the differential gravity pull of the host halo computed at $r_g$. As the subhalo's orbit is not exactly circular,
the tidal force is strongest at its pericenter, so the concentration is determined by the tidal forces at its pericenter \cite{Tormen:1997ik}:
\begin{equation}
c_{tr}(r_{p})=\frac{2\overline{\rho_h}(<r_{p})}{\rho_{crit}}-\frac{3\rho_h(r_{p})}{\rho_{crit}}\;\label{eq:ctr}, 
\end{equation}
where $\rho_h$ refers to the VLII host density profile.
We then refine the previous estimate by taking into account the pericenter radius distribution for subhalos, $\mathrm{d}P/\mathrm{d}r_p$.
Subhalos initially with $c_{t}=\Delta_0>c_{tr}\left(r_{p}\right)$ are left intact, whereas subhalos initially with $c_{t}<c_{tr}\left(r_{p}\right)$ are 
tidally stripped until $c_{tr}\left(r_{p}\right)$ is reached, e.g.: 
\begin{equation}
c_{t}'=\max\left[c_{tr}\left(r_{p}\right),\Delta_0\right]\;.\label{eq:ct1}
\end{equation}
 It is a fair approximation that the scale $\rho_{s}$ and $r_{s}$ do not change in this process \cite{Kazantzidis:2003hb}. 
By following these steps; a subhalo's mass, concentration and luminosity after tidal stripping 
are completely determined by its pericenter, mass and concentration before tidal stripping.
We then calculate the average mass, concentration and luminosity as a function of galactocentric radius, after tidal stripping and compare them against the VLII simulation. 
We find that luminosity and mass can be fitted very well, with best fit parameters $\left(m_{cut},a\right)=\left(10^{6.5},1.9\right)$. 
The $m_{cut}$ here is the minimum subhalo mass \emph{before} tidal stripping.
Our model also fits the tidal concentration very well near the host's center, although slightly deviates near its virial radius. 
This might indicate that our treatment of dynamical effects is oversimplified. Especially, 
the minimum concentration calculated by Rochi criteria is only achieved after several pericentric passages \cite{Hayashi:2002qv,Gan:2010ax,Kazantzidis:2003hb}, 
whereas in our model, we assume that all of them are already above minimum concentration.
However, the procedure is validated by very good fits to the luminosities.
Besides, it is conceptually simple and can be easily generalized to other host and subhalo mass profiles and redshifts.

Regarding the subhalos number density profile, we consider both the unbiased and anti-biased cases.
As mentioned in the section~\ref{sec:VLII}, the strong tidal force near the GC 
could make the subhalos spatial distribution to be anti-biased with respect to the host density profile. 
Indeed, for VLII itself, we find that the deviation from unbiased, starts to happen around $\sim\unit[30]{kpc}$.
However, we find that the subhalo spatial distribution for VLII is more unbiased 
with respect to VLI \cite{Diemand:2006ik,VLIIdata}, which has a lower mass resolution. 
This might indicate that the anti-bias is a result of numerical effect. The strong tidal force could 
strip subhalos until they are below or near the resolution limit, hence undetectable as subhalos.
Also, VLII subhalos which are selected by mass, show a 
more anti-biased tendency than the ones which are selected by maximum
circular velocity, confirming \cite{Diemand:2007qr,Diemand:2008in,Diemand:2009bm}. 
Keeping these in mind, we take the unbiased distribution for subhalos number density defined as: 
\begin{equation}
n_{sub}(r)=\unit[5.84]{kpc^{-3}}\left(\frac{m_{cut}}{10^{6.5}M_{\odot}}\right)^{-a+1}\frac{\rho_h\left(r\right)}{M_{\odot}\mathrm{pc^{-3}}}\;.
\end{equation}
 With this normalization, the total mass and number of subhalos with tidal mass bigger than $10^{6.5}$ 
between $\sim\unit[30]{kpc}$ and $r_h^{vir}$ in VLII simulation
is within 5\% from the value calculated with our model using best fit parameters.

The Aquarius simulation Aq-A-1~\cite{Springel:2008cc} is another simulation which has parameters similar to VLII. 
Specifically, it has nominal mass resolution, $m_p=1.71\times10^3M_{\odot}$, 
host halo mass, $M_{50}=2.523\times10^{12}M_{\odot}$, and host halo radius, $r_{50}=\unit[433]{kpc}$. 
However, their number of subhalos with mass bigger than $\sim\unit[10^7]{M_{\odot}}$ is approximately twice VLII's. 
We discuss the modification to our results in Table~\ref{tab:SubhaloContribution} and Table~\ref{tab:SubhaloDistribution} 
when we use the Aquarius normalization.

From this procedure, we can find the number density of subhalos per unit luminosity after tidal stripping,
$\mathrm{d}n_{sub}/\mathrm{d}L$. The number of photons that we receive from a single subhalo with luminosity $L$ 
and line of sight (los) distance $\lambda$ from us, for channels $ch=\gamma\gamma$ or $\gamma Z$ is given by eq.~\eqref{eq:numgam}.
By folding eq.~\eqref{eq:numgam} with $\mathrm{d}n_{sub}/\mathrm{d}L$, we can obtain the values in 
Tables~\ref{tab:SubhaloContribution} and~\ref{tab:SubhaloDistribution}. On the other hand, the contribution
to the host's $\langle\rho^2\rangle$ from subhalos is given by:
\begin{equation}
 \langle\rho_{sub}^2\rangle=n_{sub}\times \langle L\rangle\;,\label{eq:rhosbsqrd}
\end{equation}
where $\langle\ L \rangle$ is the average luminosity of the entire subhalos population after tidal stripping.

The anti-biased distribution (where $n_{sub}(r)$ is less concentrated than $\rho_h(r)$) is taken from appendix A of \cite{Kuhlen:2008aw}. 
The normalization is such that the total mass in subhalos with masses between $10^{-5}M_h$
and $10^{-2}M_h$ is $3.4\%$ of $M_h$. In the original paper, the authors normalize to $10\%$ of $M_h$. 
However, if we only include \emph{subhalos within the virial radius}, the mass fraction is only $3.4\%$. 
In this model, the authors encapsulate the tidal force from the host halo by adding 
radial dependence to the $m_{vir}-c_{vir}$ relation, so that  a subhalo closer to the GC has a higher concentration on average.

As a further refinement, we also add the Galactic disk to the VLII halo. For the Galactic disk model, we follow \cite{Catena:2009mf},
for NFW parameter and our distance to the GC, of $R_{\odot}=\unit[8.5]{kpc}$. We \emph{spherically average the Galactic disk}. The relative
difference in density between the averaged version and the original version is substantial only in the disk plane.
There is no noticeable difference to the pericenter PDF and the only modification
to unbiased distribution is that we replace $\rho_h\rightarrow\rho_h+\rho_{disk}$ in \eqref{eq:ctr}.
For the anti-biased one, the effect of host tidal forces have been taken into account by adding 
radial dependence to the $c_{vir}-m_{vir}$ relation. Therefore, in eq. \eqref{eq:ctr}, we replace $\rho_h\rightarrow\rho_{disk}$. 
We also use the same pericenter distribution as in the unbiased case.
\end{appendix}

\bibliography{2gammaLinesDD}
\bibliographystyle{apsrev}

\end{document}